%% file: main.tex
\documentclass[fleqn,10pt]{wlscirep}

\usepackage[utf8]{inputenc}
\usepackage[T1]{fontenc}

\usepackage{times}
\usepackage{epsfig}
\usepackage{graphicx}
\usepackage{amsmath}
\usepackage{amssymb}
\usepackage{soul,xcolor}        
\usepackage{dirtytalk}          
\usepackage{soul}               
\usepackage{url}

\title{Effects of auditory distance cues and reverberation on spatial perception and listening strategies}

\author[1,*]{Fulvio Missoni}
\author[2]{Katarina Poole}
\author[2,+]{Lorenzo Picinali}
\author[1,+]{Andrea Canessa}

\affil[1]{University of Genoa, DIBRIS, Genoa, 16145, Italy}
\affil[2]{Dyson School of
Design Engineering, Imperial College London, London, United Kingdom}

\affil[*]{fulvio.missoni@edu.unige.it}

\affil[+]{these authors contributed equally to this work}

\keywords{}

\begin{abstract}
Spatial hearing, the brain's ability to use auditory cues to identify the origin of sounds, is crucial for everyday listening. While simplified paradigms have advanced the understanding of spatial hearing, their lack of ecological validity limits their applicability to real-life conditions. This study aims to address this gap by investigating the effects of listener movement, reverberation, and distance on localisation accuracy in a more ecologically valid context. Participants performed active localisation tasks with no specific instructions on listening strategy, in either anechoic or reverberant conditions. The results indicate that the head movements were more frequent in reverberant environments, suggesting an adaptive strategy to mitigate uncertainty in binaural cues due to reverberation. While distance did not affect the listening strategy, it influenced the localisation performance. Our outcomes suggest that listening behaviour is adapted depending on the current acoustic conditions to support an effective perception of the space.

\end{abstract}

\begin{document}

\flushbottom
\maketitle

%
%
\thispagestyle{empty}

\section*{Introduction}
    \input{sections/Introduction}

\section*{Methods}
    \input{sections/Materials}

\section*{Results}
    \input{sections/Results}

\section*{Discussion and conclusions}
    \input{sections/Conclusion}

\bibliography{biblio}


\section*{Data availability}
The data generated and used in this study will be made available from the corresponding author upon request.

\section*{Acknowledgements}
The work by Andrea Canessa was carried out within the framework of the project HubLife Science – Digital Health (LSH-DH) PNC-E3-2022-23683267 - Progetto DHEAL-COM – CUP:D33C22001980001, founded by Ministero della Salute within ”Piano Nazionale Complementare al PNRR Ecosistema Innovativo della Salute - Codice univoco investimento: PNC-E.3”. This publication reflects only the authors’ view and the Italian Ministry of Health is not responsible for any use that may be made of the information it contains.
The authors thank Rapolas Daugintis for his valuable feedback regarding the numerical analysis.

\section*{Author information}

\subsection*{Authors and affiliations}
\textbf{Department of Informatics, Bioengineering, Robotics and Systems Engineering, University of Genoa, Genoa, Italy}
\textbf{Dyson School of
Design Engineering, Imperial College London, London, United Kingdom}
\newline
Fulvio Missoni, Katarina Poole, Lorenzo Picinali \& Andrea Canessa

\subsection*{Contributions}
All authors conceived the research together with all experiments. F.M. and A.C. implemented the framework, acquired the data and achieved the analysis. K.P. assisted in the analysis, interpretation of results and provided conceptual discussion. L.P. and A.C. assisted in the interpretation of results, provided conceptual discussion, and supervised the study. The first draft of the manuscript was written by F.M., who revised it together with L.P., A.C. and K.P. Financial support for the project was provided by A.C.

\subsection*{Corresponding author}
Correspondence to Fulvio Missoni at fulvio.missoni@edu.unige.it.

\section*{Additional information}

\subsection*{Competing interests}
The authors declare no competing interests.

\end{document}


\flushbottom
\maketitle
\begin{table}[h!]
\centering
 \begin{tabular}{|c | c | c|} 
 \hline
  & $p$ & $\lambda$ \\ [0.5ex] 
 \hline\hline
 Lateral precision & 0.2101 & 0.18 \\ 
 Lateral accuracy & 0.3285 & 0.377\\
 Polar precision & 0.0731 & 1.05 \\
 Polar accuracy & 0.0023 & 1 \\
 Quadrant error & 0.7544 & 0.7671 \\ [1ex] 
 \hline
 \end{tabular}
\caption{Statistical significance of normality test and estimated $\lambda$ coefficients of Box-Cox transformation, for the localisation performance metrics (lateral precision, lateral accuracy, polar precision, polar accuracy and quadrant error).}
\label{table:normality_check}
\end{table}
\vspace{3cm}

\begin{table}[h!]
    \centering
    \begin{tabular}{|>{\centering\arraybackslash}m{4cm}|>{\centering\arraybackslash}m{1.5cm}|>{\centering\arraybackslash}m{2.5cm}|>{\centering\arraybackslash}m{1cm}|>{\centering\arraybackslash}m{1cm}|>{\centering\arraybackslash}m{1.5cm}|}
        \hline
         Fixed Effects & Estimate & CI $95\%$ & df & t & Pr(<|t|) \\ [1ex] \hline \hline

         (Intercept) & 49.16 & [43.56, 54.76] & 67 & 17.53 & <0.001 \\ [1ex]
        Reverberation & -14.20 & [-22.79, -5.62] & 67 & -3.30 & 0.002 \\ [1ex]
        ROM & -0.48 & [-0.68, -0.28] & 67 & -4.69 & <0.001 \\ [1ex] 
        Reverberation*ROM & 0.24 & [4.56e-03, 0.47] & 67 & 2.04 & 0.046 \\ [1ex] \hline
    \end{tabular}
    
    \caption{Summary table of fixed effects estimates of the LME. The estimated coefficient (intercept and 95$\%$ Confidence Intervals (CI)) and related significance values are shown. 95$\%$ CIs and p-values were computed using a Wald t-distribution approximation.}
    \label{tab:fixed_effects}
\end{table}
\vspace{2cm}

%% file: sections/Introduction.tex
The ability of our brain to use spatial auditory cues to identify the origin of a sound is known as spatial hearing \cite{moore_introduction_2012} and is a fundamental task that supports everyday listening. Despite the challenges posed by the varying environmental characteristics and the complex three-dimensional nature of sound, the continuous integration of inputs across senses allows our brain to reduce uncertainty and facilitate adaptive interaction with the world around us. 
Spatial hearing mechanisms rely on the filtering properties of the upper body (torso, shoulders, head and ears), which cause a change in the spectral and temporal content of auditory signals as a function of the source direction.
This spatial-dependent filtering is described by the head-related transfer function (HRTF) \cite{moore_introduction_2012} and is used by our auditory system to perform spatial processing of the auditory scene. An HRTF describes both binaural (arising from the comparison of the signal at the two ears) and monaural (spectral cues originating from the ears, head and torso) auditory cues. Binaural cues (i.e., Interaural Level Difference, ILD, Interaural Time Difference, ITD and IPD Interaural Phase Difference) are used mainly for horizontal localisation judgments, whilst monaural cues support elevation and front-back judgments, where the information provided by the interaural cues is ambiguous \cite{wightman_monaural_1997}. 
However, even though simplified paradigms have allowed researchers to understand the role of the factors underlying spatial hearing mechanisms, contributing to developing intervention methods and assistive devices for hearing deficits \cite{cord_disparity_2007}, the lack of realism in such approaches (i.e., \textit{ecological validity}) has been pointed out as an important limitation \cite{keidser_quest_2020}. In recent years, more ecologically valid paradigms have gained increased interest in hearing science, highlighting the importance of investigating hearing in real-life conditions \cite{picinali_vrar_2023}. Understanding how we behave in such a complex environment is crucial to developing rehabilitation strategies effective in everyday listening tasks.

Listening is an active and dynamic process \cite{carlile_perception_2016} and the continuous integration of dynamic auditory cues with visual \cite{alais_ventriloquist_2004} and proprioceptive \cite{brimijoin_contribution_2013, vliegen_dynamic_2004} signals has an important role in our perception \cite{wallach_role_1940}. Previous studies have shown that, in anechoic conditions, the use of head movements coupled with head-tracking systems \cite{wightman_resolution_1999, hendrickx_influence_2017} can facilitate the accurate localisation of free-field (i.e., loudspeakers) \cite{perrett_effect_1997, gaveau_benefits_2022, iwaya_effects_2003, kato_effect_2003} and binaural (i.e., rendered through a pair of headphones) spatial audio. In particular, when head movements are allowed, front-back confusions are significantly reduced both, during free-field \cite{brimijoin_contribution_2013, macpherson_head_2011} and binaural presentations \cite{perrett_effect_1997, wightman_resolution_1999}. Moreover, recent works have shown that the use of active interaction with head movements is an effective rehabilitation strategy when hearing is impaired \cite{valzolgher_reaching_2023, gessa_spontaneous_2022} by using proprioceptive information to update the sensory information in a contingent manner. Optimal head orientation is also beneficial when trying to understand speech in a noisy environment \cite{grange_benefit_2016, grange_turn_2018}, allowing a better separation between target and masking signals. Finally, even though the use of movements is advantageous in different listening activities, these studies have found significant individual variations in head movements when performed without specific instructions. Despite the large differences, head rotational motion appears to be a spontaneous behaviour generally devoted to exploring the surroundings, without a specific strategy. However, the factors that drive these spontaneous behavior are still unknown.

Beyond anechoic conditions, which are typical of lab environments and very rarely encountered in real life, it is well established that environmental acoustic cues are essential for the perception of space in static conditions. However, little is known about how these factors affect our dynamic listening.
The sound waves emitted by an auditory source are partially reflected and scattered by nearby objects and surfaces, creating a more diffuse sound field, known as reverberation \cite{zahorik_spatial_2021}. As the distance increases the energy ratio between the direct and diffuse signals decreases, providing a strong cue for distance judgments. This relationship, in combination with other auditory cues such as the relative sound level and the frequency content of the auditory source (see \cite{kolarik_auditory_2015} for a review), is used when estimating the distance of a sound source. On the other hand, previous evidence have shown that, in stationary listening, the interference of the reflective signals with the direct sound disrupts the coherence of the signals between the two ears \cite{beranek_concert_2010}, reducing the available binaural cues and thus lowering the localisation accuracy in the horizontal plane \cite{devore_accurate_2009,giguere_sound_1993}. To maintain accurate localisation in a reverberant environment, it has been suggested that by adjusting the relative perceptual weights of ILD and ITD cues \cite{ihlefeld_effect_2011} our brain is able to build a robust encoding of spatial cues. For example, evidence has been provided to show that, as the reverb level is increased, our brain relies more on ILD than ITD \cite{devore_accurate_2009}. Nonetheless, the effect of reverberation in active listening is still under-investigated.

In a key study carried out more than twenty years ago, Begault and colleagues \cite{begault_direct_2001} investigated the impact of reverberation and head-tracking, and their interaction, on spatial perception of a virtual sound source. Nine naive listeners had to localise short speech stimuli (3s) positioned in the horizontal plane at six different azimuth directions (0$^\circ$, $\pm45^\circ$, $\pm135^\circ$, 180$^\circ$). This task was repeated with three levels of reverberation (anechoic, early reverberant and full reverberant) and with and without head-tracking. Participants were asked to move their heads only in the head-tracked condition, without any restrictions. The response was given after each stimulus presentation via a computer mouse using an interactive graphic on a monitor screen. Their outcomes showed individual effects of both reverberation and head-tracking, in localisation perception without any interaction between the two factors. The reverberation level decreased the horizontal error to the detriment of elevation precision. The presence of head-tracking had a significant effect on the front-back confusions, reducing them by a ratio of about 2:1, but the localisation precision of within-quadrant stimuli was not affected, suggesting that head movements were mainly used for identifying the correct quadrant rather than to improve the location response within it. This study was fundamental to better comprehending the interaction between head movements and reverberation in spatial auditory perception, even though the paradigm used hardly represented the listening conditions usually encountered in the natural world. For example, the participants were asked to move continuously during listening, and only stimuli at a fixed distance were investigated, therefore not considering the fact that source distance can, directly and indirectly, affect our perception of space.

In summary, spatial hearing is a complex multi-modal task in which the integration of self-generated dynamic cues with visual and motor information is fundamental. In addition, the presence of reverberation has often a detrimental effect on binaural localisation cues, impacting our localisation performance in static localisation. However, there are little evidence on how the acoustic condition (i.e., the presence of reflections or not) affects our listening in dynamic conditions. The work of Begault et al. (2001) \cite{begault_direct_2001} suggests that the presence of reverb does not affect the perception of space with head-tracking, although it is reasonable to argue that our brain adapts the listening strategy depending on the acoustic context, and the use of the active movements might be used to reduce the uncertainty of uncorrelated binaural cues in presence of reverberation.  

This study takes a more ecologically-valid approach to studying spatial hearing performance. Here we aim to understand the effects of listener movement, the acoustic condition and distance on localisation accuracy in the horizontal (also known as lateral) and vertical (also known as polar) directions. More specifically, we firstly look at the role of head rotations in different acoustic conditions in spatial perception during an active listening task, investigating whether or not the presence of reverberation has an impact on the listening strategy. Then we investigate how the target distance indirectly affects the auditory spatial perception in the lateral and polar dimensions. 

Our hypotheses are:
\begin{enumerate}
    \item Reverberation negatively affects localisation performance. Due to the diffuse nature of the reverberant signals the presence of reverberation results in a reduction of binaural localisation cues. This is expected to negatively affect localisation performances in the lateral dimension, at least for faraway sources, i.e. where the reflected component is more present if compared with the direct one.
    \item The use of head movements is increased in the presence of reverberation. Due to the reduction in the reliability of the localisation cues, when reverberation is present head movements become more important (and are therefore used more frequently) for the localisation of the sound sources.
\end{enumerate}

%% file: sections/Materials.tex
\paragraph{Participants}
For this study, we recruited a total of 26 self-reported normal hearing participants (19 female, 7 male; age: mean = 26, SD = 3), however, one was excluded due to a lack of understanding of the experiment. All subjects self-reported a low level of experience with binaural audio assessments. To remove any order effects, and minimise exposure to binaural spatial audio, they were randomly divided into two groups to test a specific acoustic condition either: an anechoic chamber (n = 13 [10 female]; age: [24; 30]) or a reverberant room (n = 13 [9 female]; age: [24; 30]).
This study was approved by the University of Genoa Research Ethics Committee and performed following relevant guidelines and regulations. Participants signed a consent form to take part in the experiment, in which they were informed about the experiment and that they may withdraw from the study at any time.

\paragraph{Equipment and Experimental setup}
The participants wore a head-mounted display (HMD) (HTC Vive Pro 2; HTC Vive system, HTC Corporation, New Taipei City, Taiwan) and a pair of circumaural headphones (Sennheiser 280 PRO). A handheld Vive Controller was used to interact inside the virtual scene. The VR system is tracked in real-time in six degrees of freedom (DoF) through the HMD tracking system, using two base stations (SteamVR Base Station 2.0) positioned at two corners within the room, providing a submillimeter spatial precision \cite{verdelet_assessing_2019}. This apparatus was connected to an Alienware Aurora Ryzen Edition computer running a 64-bit Windows 11, equipped with a 12-cores AMD Ryzen 9-5900, 32GB of RAM and a graphic card with dedicated RAM of 10GB (GeForce RTX 3080, NVIDIA Inc.), controlled using Steam VR 2.0 and Unity 3D (Unity Technologies, San Francisco, CA). 
The acoustic scenes were rendered in real-time using the Unity Wrapper module of the 3DTune-In Toolkit \cite{cuevas-rodriguez_3d_2019} in Unity (version 2021.3.21f), using the head tracking data of the HMD.

\begin{figure}[h!]
    \centering
    \includegraphics{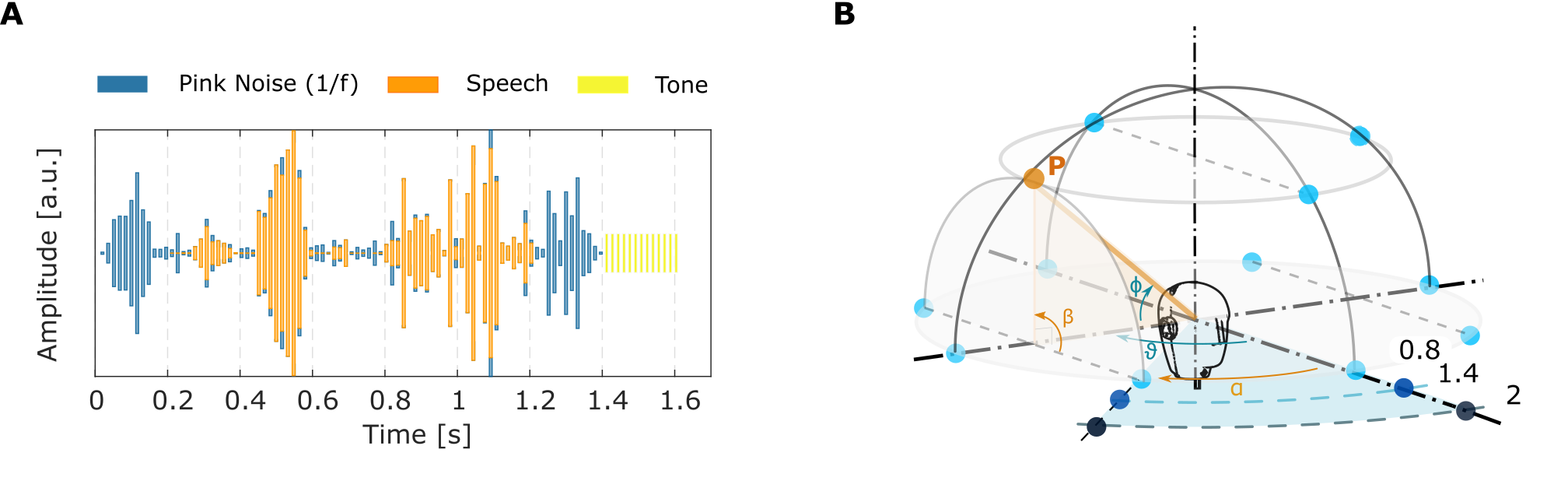}
    \caption{\textbf{Experimental paradigm.} A) Representation of the used auditory stimulus which is a sum of three components: pink noise, speech and a pure tone; B) Representation of coordinate systems and stimuli locations represented with both, the spherical (in orange) and the interaural coordinate systems. $\theta$ and $\phi$ are respectively the azimuth and elevation, while $\alpha$ and $\beta$ represent the lateral and the the polar angle. For instance, point P is in (90$^\circ$, 45$^\circ$) in the spherical coordinate system and (45$^\circ$, 90$^\circ$) the interaural coordinates. Dots show the tested directions, which are repeated for three different distances (expressed in meters): 0.8, 1.4 and 2.}
    \label{fig:methodology}
\end{figure}

\paragraph{Stimuli}
In order to ensure that the audio stimuli contained rich cues for localisation, these were generated following the same procedure proposed in \cite{steadman_short-term_2019}. Each stimulus comprised a combination of two 200ms lasting segments of pink (1/f) noise at the beginning and at the end of a 1s segment of speech (female talker, randomly selected from the TSP speech database \cite{kabal_tsp_2008}) and a final segments of $1$kHz tone ($200$ms) (see Fig.\ref{fig:methodology}A). The amplification was calibrated in order to generate $64$dBA at $2$m. The total duration of the stimulus (1.6s) was chosen to prevent participants from moving too easily, which could happen with longer stimuli, but still allowing the movement.

Stimuli, differing across experimental phase  (see \textit{Experimental procedure}), were presented over headphones and rendered using two different HRTFs selected from the IRCAM LISTEN database \cite{noauthor_ircam_nodate}. Each HRTF was used for a different experimental phase to avoid any learning effects. Among the available HRTFs, we randomly selected those corresponding to the participant numbers IRC0008 and IRC0013.
Two acoustical conditions were considered. The first consisted of a pure anechoic room where only the direct sound was reproduced. The second condition was a reverberant one, obtained using the Reverberant Virtual Loudspeaker (RVL) method \cite{engel_assessing_2022}. Table \ref{table:room_reflections} shows the parameters of room reflections.
In this manuscript, When it is not specified, we used the term "acoustical condition" to refer to different reverberation levels.

\begin{table}[h!]
\centering
 \begin{tabular}{ c | c | c | c| c | c | c| c | c | c| c | c } 
 \hline
   Frq.band [Hz] & 31.5 & 63 & 125 & 250 & 500 & 1k & 2k & 4k & 8k & 16k & Broadband \\
 \hline\hline

 T30 [s] & 0.787 & 0.799 & 0.794 & 0.637 & 0.561 & 0.834 & 0.707 & 0.672 & 0.549 & 0.376 & 0.581 \\ 
 EDT [s] & 0.840 & 0.699 & 0.747 & 0.607 & 0.653 & 0.663 & 0.675 & 0.623 & 0.530 & 0.372 & 0.594\\ 
 \hline
 \end{tabular}
\caption{\textbf{Room reflections parameters.} Reverberation time (T30) and early decay time (EDT) for octave frequency bands (in Hz). The last column shows the broadband values.}
\label{table:room_reflections}
\end{table}
The distance attenuation followed an exponential rule that reduced the sound level every double distance. The computation was independent for the direct ($-6$dB) anechoic) and reverberant ($-3$dB) sound path to emulate what happens in real conditions \cite{cuevas-rodriguez_3d_2019, chowning_simulation_1977}. Sources were modified with near-field correction filters to take into account the spherical nature of the sound wave when the source is close to the listener (see \cite{cuevas-rodriguez_3d_2019} for more details). All stimuli were generated and stored in $48$kHz, $16$-bit format.

\paragraph{Experimental procedure}\label{par:exp_proc}

Participants sat on a chair which allowed them to rotate, in a quiet (average 40dB SPL, A-weighted), empty room (width 3m, depth 4m, height 3m), wearing the headphones and the HMD to do the experiment. The rendered virtual environment was a rectangular-shaped room (width 7m, depth 7m, height 3m) with a black fixation cross that marked the front of the room. In the room we put $12$ visible virtual loudspeakers at $1.5$m from the listener, uniformly distributed in the upper hemisphere (see Fig.\ref{fig:methodology}B). The horizontal plane (elevation equal to $0^{\circ}$) was sampled with eight azimuth values with steps of $45^{\circ}$ and the elevated plane of $45^{\circ}$ with four orientations spaced by $90^{\circ}$. For each subject, the relative position of the virtual speakers was pseudo-randomly jittered 20cm along each cartesian axis in order to have different location arrangements between subjects.
The protocol consisted of two phases: a familiarisation phase and the subsequent experimental phase.
In the familiarisation phase, listeners were allowed to use the controller to autonomously select a loudspeaker, making it play a stimulus. In this phase, we decided to render the stimulus through the IRC0008 HRTF. Subjects were instructed to try out all the speakers but to select only one loudspeaker at a time. Once a loudspeaker was selected, subjects were allowed to play it multiple times. In this way subjects familiarised themselves with the virtual sound environment, the binaural spatial sounds originating from different locations, the auditory source and the controllers used during the experiment. Visual and auditory stimuli were collocated and fixed in the VR world reference system to allow each participant to freely move their head. 
Whenever participants felt comfortable they could move to the subsequent experimental phase.
In the subsequent block, the speakers disappeared from the room and the test started. The task was an iterative repetition of the following steps: to point the head gaze forward to the fixation cross; to start a new trial by pressing the controller button; to report the stimulus location by pointing the head gaze towards the perceived direction. 
A trial only began once the listeners had aligned their head to face forward by using a tiny red sphere, which indicated the head gaze, within a 16$^\circ$ circular area located in front of them.
We tested each of the $12$ positions for $5$ times at three different distances$-$at  $0.8$, $1.4$ and $2$m$-$for a total of $12\times3\times5 = 180$ trials for each subject.
In each trial, we rendered the stimulus through the IRC0013 HRTF, delayed by $1.7$s ($\pm0.5$s) with respect to the start to ensure the steady position of the head, and with a position jitter of $1$cm along each cartesian axes.
It's worth noting that, head movements were allowed when the stimulus was played but subjects were unaware of this possibility. To investigate the role of spontaneous head movements, they were instructed to use the strategy that they preferred.

\paragraph{Localisation and head movements metrics}

Subject head tracking data (i.e., HMD position and rotation, sample rate: $90$Hz) were acquired during the entire duration of a trial to assess the listening strategy during the task execution. To assess localisation performance, the subject's response location was recorded as the instantaneous orientation of the HMD when the participants pressed a button to confirm their choice.

The target's position and head's orientation were described in the interaural coordinate system \cite{morimoto_localization_1984}, with lateral and polar angles that directly map coordinates and localisation cues. The lateral angle was the angle between the frontal axis and the sound source position vector and ranged from $-90^\circ$ (right) to $90^\circ$ (left). Being related to the lateral position, this coordinate corresponds to the mechanism for the horizontal perception. The polar angle was the elevation angle within the sagittal plane between the horizontal plane (or interaural plane) and the source position. It is related to the front-back and vertical perception. The polar angle ranged from $-90^\circ$ (front, below eye level) to $270^\circ$ (rear, below eye level). 

Employing this coordinate system we assessed the individual localisation performance by computing the precision (i.e. standard deviation) and unsigned accuracy of the localisation errors in the lateral and polar dimension \cite{middlebrooks_virtual_1999}, defined as the absolute difference between the response (i.e., head orientation) and target position. The polar error was computed after removing all the front-back confused trials, which were measured by the quadrant error rate. Quadrant error rate was computed as the percentage of responses in which the weighted polar error exceeded $\pm45^\circ$ considering all targets within lateral angles of $\pm60^\circ$, as suggested in \cite{majdak_3-d_2010}.
We decided to characterize the head movement of each subject in terms of the range of motion (ROM) and movement onset (MO). To estimate the ROM for each subject we first, computed the Yaw angle Maximum Absolute Deviation (MAD) across time for each trial and then averaged the MAD values across trials. 
To compute MO, we first calculated head yaw angular velocity, and we removed noise with a Savitzsky–Golay finite impulse response filter. A speed threshold is then applied to define head movement onset. The threshold was set at the $5\%$ of the peak velocity. To obtain the onset time of a head movement, the algorithm walked backwards in time from each velocity peak over threshold and marked movement onset time as the last point where the head’s angular velocity was above threshold.
We decided to use the statistical parametric mapping method (SPM1D) \cite{pataky_generalized_2010}, which is specifically designed for continuous field analysis, to perform a one-dimensional t-test contrasting the time profiles of the yaw angles obtained in the two acoustic conditions.

\paragraph{Behavioral data analysis}
 
We used the Anderson-Darling test (MatLab\texttrademark, Mathworks) to check the normality of the data, and in case when the normality was not respected data were corrected with Box-Cox transformation. 
A two-way repeated measures analysis of variance (ANOVA) was conducted to assess the effect of the acoustic condition and the target distance, on sound localisation error. The between- and within-subjects factors were acoustic condition and the target distance respectively.
Moreover, a linear mixed effect model was used (estimated using ML and nloptwrap optimizer) to investigate the effect of head rotation magnitude and acoustic condition on the quadrant error rate. ROM, the acoustic condition and the interaction term were treated as fixed effects and the subject as a random intercept. To capture the individual differences due to the different reverberation levels and target distances, both were treated as random slopes. 
The localisation performance analysis was conducted in MatLab\texttrademark  (Mathworks) using the AMT (Auditory Modelling Toolbox) version 1.2.0 and custom scripts. Statistical analysis was conducted in Jamovi and R version 4.3.2 (R Core Team, 2020). 

\paragraph{Acoustic Analysis}

Binaural signals were analysed as a function of distance and room acoustic properties (i.e., reverberation) to identify the influence of these components in spatial auditory perception. To do this, the binaural localisation cues (ITD, ILD and interaural coherence) were computed for different directions to identify how each of these cues was affected. As the majority of head movements were in the horizontal plane, the analysis was focused on horizontal localisation judgments.
All the simulations were rendered through the 3DTune-In Toolkit App using a custom Python script. The analysis was conducted in MatLab using AMT and the Binaural-SH package \cite{engel_assessing_2022}.
A white noise burst of 100ms was uniformly distributed in the interaural plane (elevation angle equal to $0^{\circ}$) along 72 directions (azimuth step of $5 ^{\circ}$) and four distances (linearly spaced from $0.5$m to $2$m). Firstly, the static listening condition is characterised with the binaural cues (ITD, ILD) and the interaural coherence. The Interaural Cross-Correlation (IACC) coefficient \cite{iso_3382-1_acoustics_2009} was used to assess the interaural coherence, which is defined as the absolute maximum of the time-limited normalized cross-correlation function between the signal at left and right ears. ITD was computed using the MaxIACCe method, as described in \cite{katz_comparative_2014}, which searches for the temporal lag corresponding to the Maximum of IACC in HRIR energy envelopes, after applying a low-pass filter (3kHz). The ILD was estimated, using a high-passed (1.5 kHz) Head-Related Impulse Response (HRIR), computing it independently for 30 equivalent rectangular bandwidths (ERB) and then computing the average as suggested by \cite{mckenzie_interaural_2019}. 

%% file: sections/Results.tex
\paragraph{Kinematic Analysis: reverberant cues encourage the use of rotational movements}

\begin{figure}[h!]
    \centering  \includegraphics[width=\textwidth]{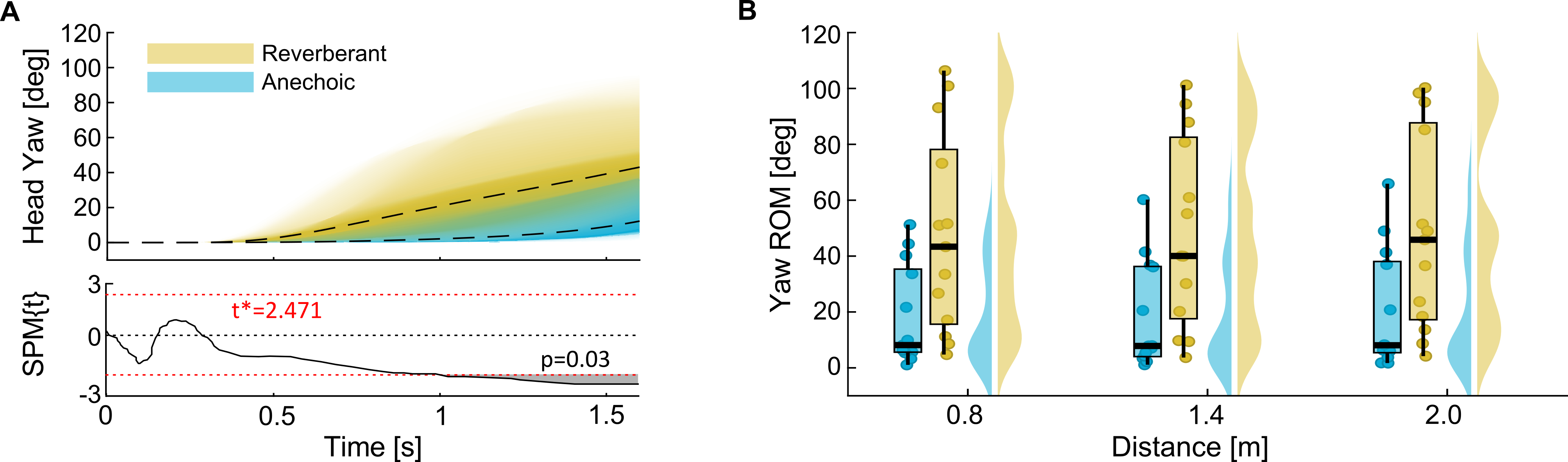}
    \caption{\textbf{Kinematic analysis}. Data are presented grouped per reverberation level (Blue: anechoic, Yellow: reverberant) A) (top) Density plot of yaw trajectories (in degrees) over time (in seconds) for each trial and participant, grouped by reverberation level; Dotted lines represent the median trajectory per group computed across trials, participants and target position; (bottom) The t‐test statistic. The critical threshold t$^*$ = 2.471 (red dashed line) is exceeded at time = 1.23s,  with a supra‐threshold cluster probability of p=0.03, indicating a significantly higher yaw amplitude angle in the reverberant condition. B) ROM (in degrees) grouped by target distance (in meters) and reverberation level}
    \label{fig:cinematic}
\end{figure}

To identify any difference between listening strategies we analysed yaw rotations in reverberant and anechoic conditions.
Looking at head rotations in the horizontal plane (i.e., yaw) shows that the two groups use different strategies. Figure \ref{fig:cinematic}A shows a density plot of the absolute value of yaw rotation over time, separated for reverberation level, during stimulus reproduction. Compared to the anechoic group, which is less engaged in active listening, the reverberant one exhibits earlier movement onset (anechoic: mean = 1.16, SD = 0.32; reverberant: mean = 0.79, SD = 0.37s; p=0.014, Wilcoxon ranksum test) and a significant difference in the amplitude of yaw movement (p=0.03, spm1d paired samples t‐test).
This suggests that the acoustic condition affects the rotational movements despite the wide distribution of the trajectories across groups. 

This difference was confirmed by the statistical analysis on ROM (see Fig. \ref{fig:cinematic}B).
An ANOVA test (normality Anderson-Darling: p<0.001) showed a main effect of the acoustic condition on ROM (F(1,23) = 5.79, p = 0.025) and no significant effect of either, distance (F(2,46) = 0.467, p = 0.630) or interaction (F(2,46) = 0.606, p = 0.550). This indicates that not the distance of the stimuli, but the acoustic-related cues affect the use of head movements.

\paragraph{Acoustical Analysis: reverb disrupts the interaural coherence}

\begin{figure}[h!]
    \centering
    \includegraphics{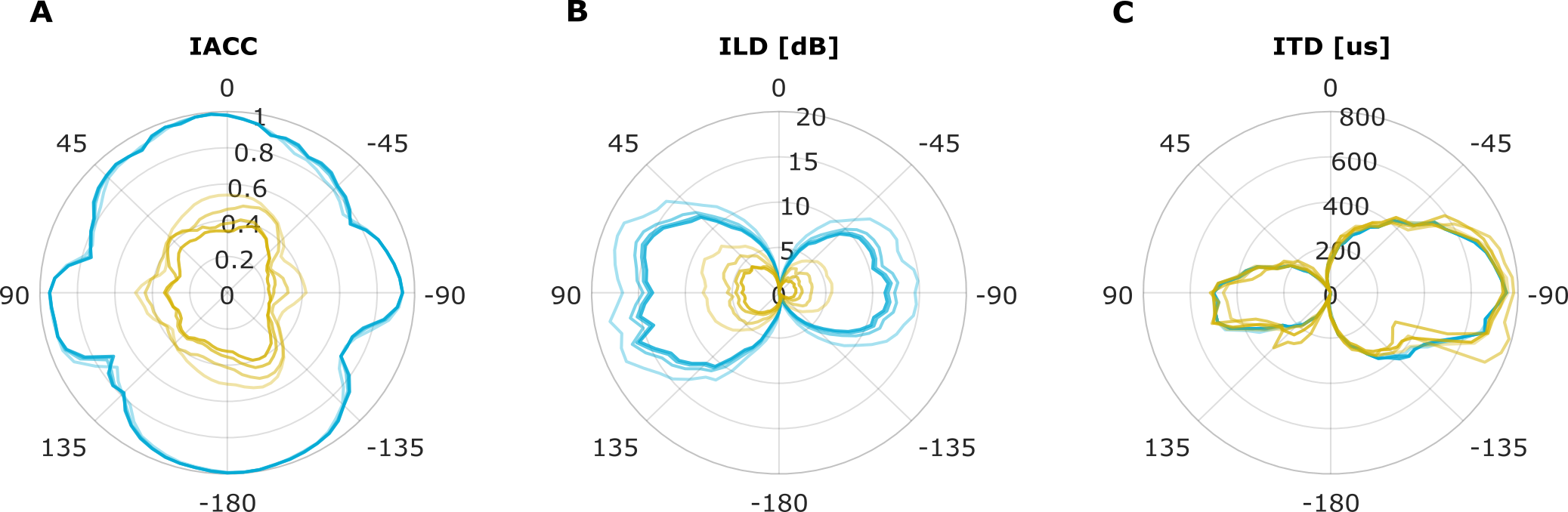}
    \caption{\textbf{Effect of distance-related cues on binaural cues.} Binaural cues as a function of target locations (angles in degrees) and reverberation level (blue: anechoic; yellow: reverberant). Shading is relative to the target distance: from brighter (near: 0.5m) to darker (far: 2m). (A) Interaural cross-correlation (B) Interaural Level-difference (in dB) (C) Interaural Time-difference (in $\mu$s)}
    \label{fig:aa_spatial_cues}
\end{figure}

As different listening strategies could be a direct result of the influence of reverberation on the localisation cues, we decided to conduct an acoustical analysis on the binaural cues as a function of reverberation and stimulus distance (see Fig.\ref{fig:aa_spatial_cues}).
In the anechoic condition, the static binaural signals are highly correlated along the entire horizontal plane (mean IACC = $0.88$), with the largest values along the cartesian axes (left-right: IACC = 0.94, and frontal-rear: IACC = 0.98) and lowest values along the diagonal orientation (azimuth = $\pm45^\circ$: IACC = [0.91, 0.8]; azimuth = $\pm135^\circ$: IACC = [0.78, 0.88]). Furthermore, there is a decreasing effect of the target's distance on the ILD which is greater only when near (0.5m) and far distances (2m) are compared. 
There is no observable effect of distance on ITD and IACC in the absence of reflections.

These results show that reverberation lowers the interaural coherence (mean IACC = $0.37$) and concurrently the available binaural cues (see Fig.\ref{fig:aa_spatial_cues}A). Reverberation also affects the level-dependent cue reducing the ILD in terms of absolute values (see Fig.\ref{fig:aa_spatial_cues}B). In contrast, ITD is more robust to reverberation, even though the presence of reflections increases the variability between target locations (see Fig.\ref{fig:aa_spatial_cues}C).
In summary, reverberation is responsible for the decrease of cross-correlation between binaural signals, reducing the available static binaural localisation cues.

\paragraph{Distance affects localisation errors but not the listening strategy}

\begin{figure}[h!]
    \centering
    \includegraphics{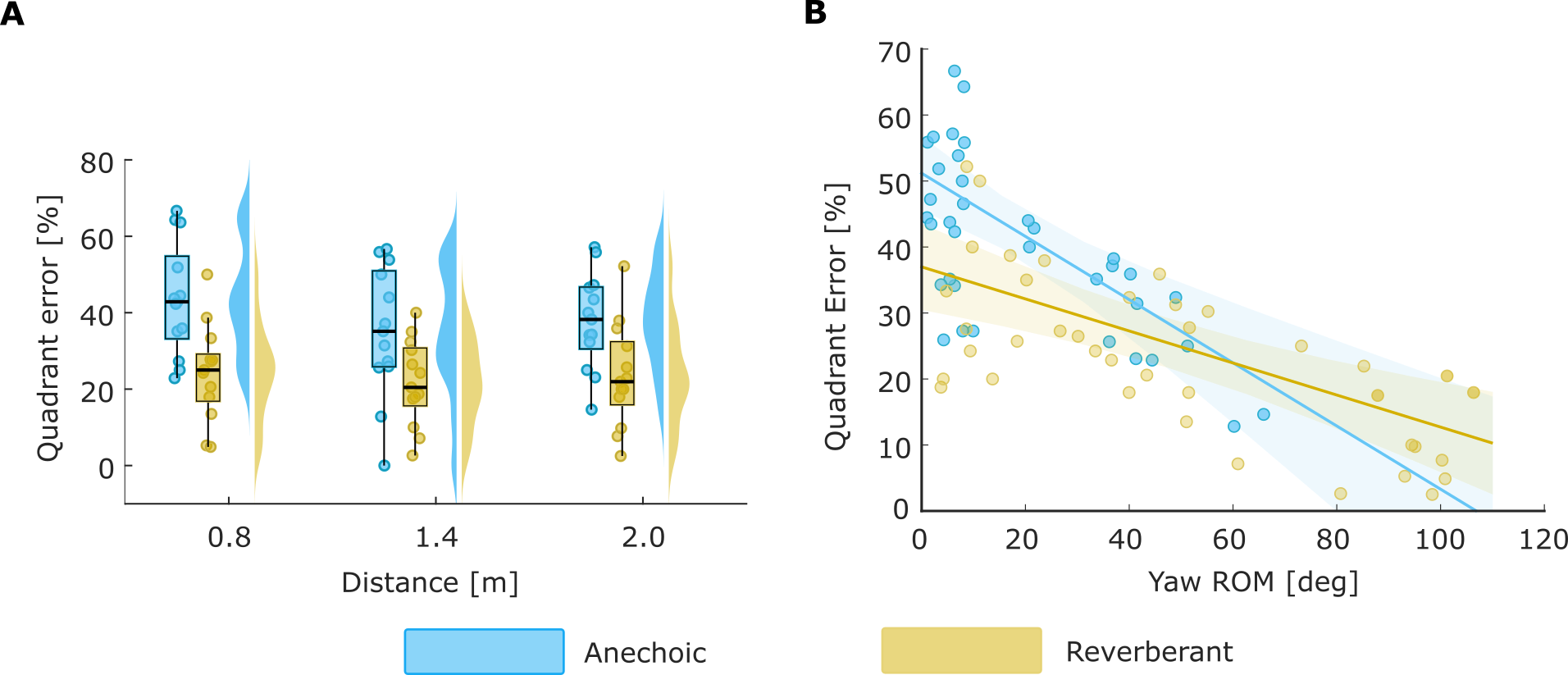}
    \caption{\textbf{Distribution of quadrant error and effect of motion.} A) Quadrant error rate as a function of target distance (in meters) and reverberation level (blue: anechoic, yellow: reverberant). B) Quadrant error rate as a function of ROM (in degrees) and reverberation level (blue: anechoic, yellow: reverberant). Scatter points represent ROM for each subject and each distance. The two lines indicate the fitter regression model. Dashed lines indicate the 95\% confidence interval of the linear regression model.}
    \label{fig:LME}
\end{figure}

Subsequently, we tested the influence of the stimulus's distance on localisation errors and head rotations to investigate whether or not distance modulation of the binaural cues was reflected in the listening behaviour or in the localisation performance.
The localisation errors grouped by target distance and acoustic condition are shown in Fig.\ref{fig:localisation_data}. 
The results of the normality test on the localisation errors are displayed in Supplementary Table S1 (see Supplementary Material), showing the non-normal distribution of the acquired data, which were corrected with Box-Cox transformation (see Supplementary Table S1 for estimated $\lambda$ factors).
We observed a main effect of distance in localisation errors, with opposite effects in lateral and polar judgments. In particular, in the lateral dimension both, precision (F(2,46) = 8.881, p<.001, see Fig.\ref{fig:localisation_data}A) and accuracy (F(2,46) = 8.355, p<.001, see Fig.\ref{fig:localisation_data}B) errors are significantly decreased in both groups. Post-hoc analysis using the Bonferroni correction shows that lateral error in the near space (0.8 m) is significantly higher than the middle (1.4 m; precision: p = 0.008; accuracy: p=0.015) and far (2 m; precision: p=0.004; accuracy: p = 0.009) space. 
On the other hand, we see a significant main effect of acoustic condition on the polar precision (F(2,46) = 6.54, p = 0.003, see Fig.\ref{fig:localisation_data}C) but not on the accuracy (F(2,46) = 2.390, p = 0.103, see Fig.\ref{fig:localisation_data}D), indicating the reliability rather the accuracy of the response is reduced for near sound sources. Pairwise comparisons (t-test, Bonferroni corrected) revealed polar precision errors in the near space (0.8 m) are significantly lower than in the far (2 m; p = 0.001) distances. Overall, these findings suggest that, as the distance increases, both lateral precision and accuracy improve but the reliability of the responses in the polar dimension decreases.
Finally, no significant effect is reported for quadrant error (F(2,46) = 1.2084, p = 0.308, see Fig.\ref{fig:LME}A).
   
\paragraph{The different listening strategy affects the quadrant errors, but not the polar localisation performance}

Finally, we looked at the difference in localisation performance to identify the effect of the different listening strategies on spatial perception.
A statistically significant effect of the acoustic condition is observed in the lateral precision (F(1,23) = 4.39, p = 0.047, see Fig.\ref{fig:localisation_data}A) but not in the accuracy error (F(1,23) = 2.90, p = 0.102, see Fig.\ref{fig:localisation_data}B). Also in the polar plane, the effect of acoustic condition is not significant (precision: F(1,23) = 0.0546, p = 0.817, see Fig.\ref{fig:localisation_data}C; accuracy: F(1,23) = 0.204, p = 0.656, see Fig.\ref{fig:localisation_data}D), indicating that the reported difference in the range of movements has a relative impact on localisation ability, most probably because the motions are free and unconstrained.

However, the quadrant error is significantly different between groups (F(1,23) = 12.4, p = 0.002, see Fig.\ref{fig:LME}A). Due to the strong relationship between motion and quadrant error, a linear mixed effect model is fitted to investigate the effect of the magnitude of the head rotation and the reverberation level on the quadrant error (see Fig.\ref{fig:LME}B).
\begin{figure}[h!]
    \centering   \includegraphics{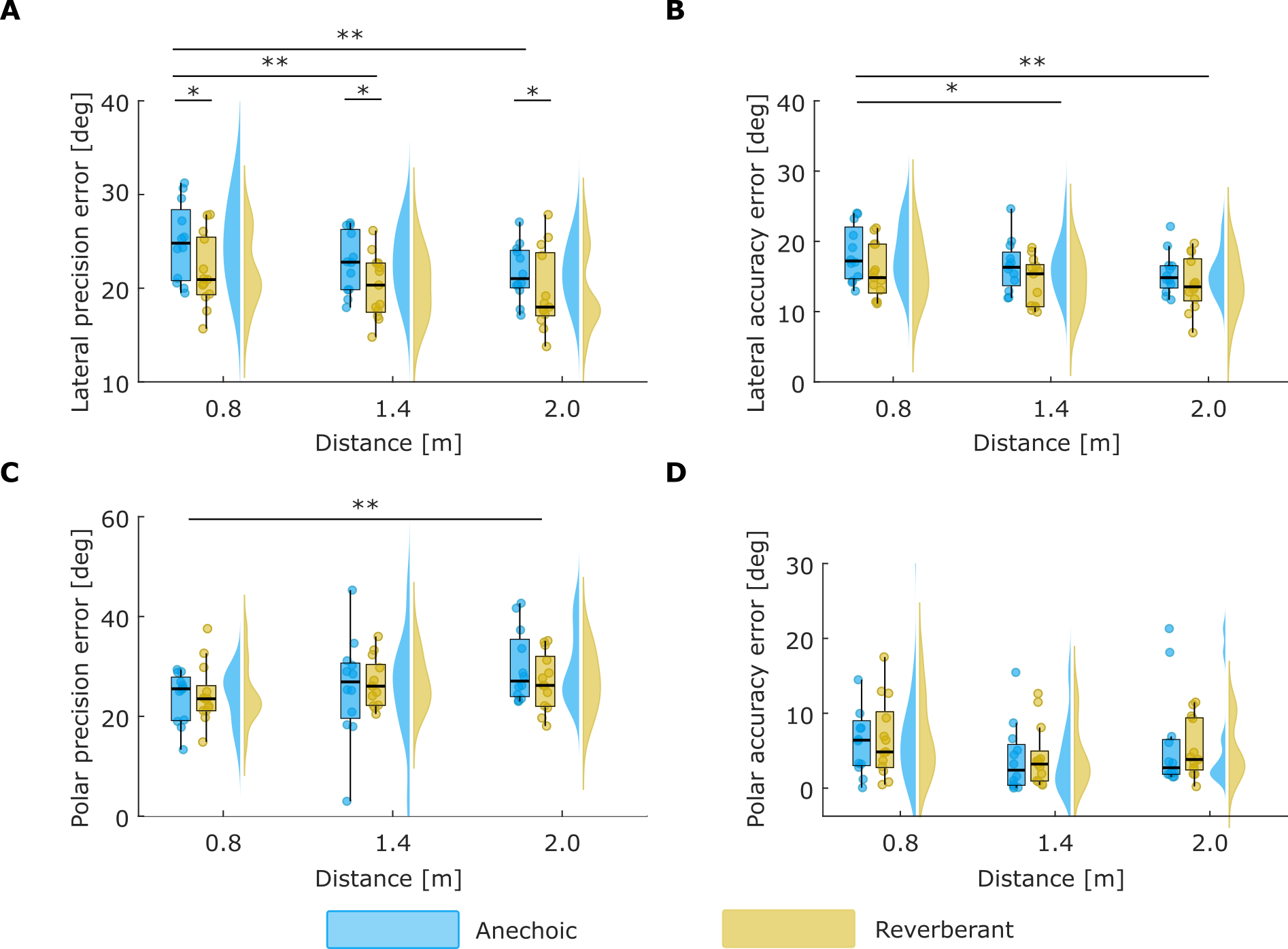}
    \caption{\textbf{Distribution of Localisation errors (in degrees).} Lateral precision (A) and accuracy (B) error, Polar precision (C) and accuracy (D) errors grouped for target distance (in meters) and reverberation level (blue: anechoic, reverberant: reverberant). Significance indicators show the outcomes of the repeated measures ANOVA.}
    \label{fig:localisation_data}
\end{figure}
The model's total explanatory power is substantial (conditional $R2$ = 0.81) and the part related to the fixed effects alone (marginal $R2$) is 0.59 (See the supplementary information for more details on the model). 
We observed a main negative effect of ROM on quadrant error rate (t(67) = -4.69, p < 0.001) confirming that increasing the range of head movement is beneficial for decreasing the quadrant error. Moreover, the quadrant error rate is negatively affected by the acoustic condition (t(67) =  -3.30, p = 0.002) and positively affected by the interaction term (t(67) = 2.04, p = 0.046) as the relative level of quadrant errors with small movements is higher in anechoic condition. This effect is most likely a consequence of the increased level of internalisation without reverberation.

%% file: sections/Conclusion.tex
In this work, we were interested in the investigation of the influence of acoustic conditions and source distance on listening strategies in ecological settings. Our objective was to understand whether and how the listening behaviour is adapted to different environmental conditions to support an effective spatial perception.
Our results suggested that the presence of reverberation and distance positively affect the localisation performance in the lateral dimension. On the contrary, while reverberation had no significant effect on polar error, distance negatively affected the polar error. Moreover, we observed an increased use of head movement in the reverberant group, suggesting that head movements are probably more important for sound localisation when the level of reverb is increased.

It is well known that the perceptual illusion of spatial sounds reproduced over headphones (or externalisation) is highly dependent on the presence of reverberation \cite{leclere_externalization_2019}. Catic and colleagues \cite{catic_effect_2013} have suggested that the temporal fluctuations due to the reflections of both, ILDs and interaural coherence, are responsible for this phenomenon. More in general, when the synthesised sounds mismatch with the prior listener's experience (see \cite{best_sound_2020}), the auditory illusions tend to collapse inside the head \cite{plenge_differences_1974,jeffress_lateralization_1961}, influencing the localisation ability.
We found that the average lateral and polar precision errors were significantly lower in the reverberant group ($20.5^\circ$, $26^\circ$)  than the anechoic one ($23.1^\circ$, $26.5^\circ$). Though in our study we grouped subjects by the reverberation levels, these values are very close ($\pm3^\circ$) to the unsigned azimuthal and elevation errors reported by Begault et al. (2001) \cite{begault_direct_2001}, where all participants were tested in both anechoic, early reverberant and late reverberant conditions. Moreover, they found that the azimuthal errors were significantly reduced when the reverberation level increased, to the detriment of elevation precision suggesting that was dependent on different levels of externalisation across conditions.
We didn't find a significant effect of acoustic condition in the polar dimension but of distance, which might be due to differences in the experimental paradigm. Here subjects were tested with sources located at different distances, whilst Begault et al., only tested a fixed distance. We found that the presence of distance affects both ILD and IACC, independently of the presence of reverberation. The gain modulation due to the changing of the source distance among trials caused the variation of the ILD, allowing a temporal oscillation of both ILD and IACC between trials and thus affecting the externalisation perception as suggested by Catic et al. 

In our familiarisation phase, participants were instructed to test different auditory positions that were visually rendered without a fixed time. Although the familiarisation time was determined by the listener, we found that the average value per subject was short (mean = 328s, SD = 158s), suggesting that it was too limited to significantly affect their localisation ability. It has been shown that listeners trained by using visual feedback (i.e., visual representation of the sound source) can adapt to a different HRTF in a short amount of time of the order of a few hours \cite{steadman_short-term_2019, majdak_3-d_2010}. Furthermore, it has been proposed that shorter training (less than thirty minutes) could improve the localisation ability \cite{mendonca_improvement_2012, parseihian_rapid_2012}, although such rapid adaptations are usually linked to the procedural learning \cite{ortiz_contributions_2009}.

In contrast to the anechoic condition, where the use of static localisation cues was prioritised, the reverberation seemed to promote the adoption of an active strategy. The earlier movement onset of the reverberant group (790 ms vs 1160 ms) indicates that the presence of reflections induced the subjects to actively localise the sound source. Moreover, we didn't find any modulation effect of the target distance in the ROM, suggesting that the use of movements was more dependent on the presence of reverberation than the source distance. This different strategy likely influenced the quadrant error rate but not the localisation performance.
The strong correlation between head movements and the reduction of front-back confusion (or reversal rate) is extensively documented with either real (i.e., loudspeakers) \cite{perrett_contribution_1997,iwaya_effects_2003,mcanally_sound_2014} or virtual sources (i.e., binaural reproduction) \cite{wightman_resolution_1999, mclachlan_dynamic_2023}. For example, McAnally and Martin (2014) \cite{mcanally_sound_2014} have shown that, when subjects are tested with sources in free-field, small movements ($\pm4^\circ$) significantly reduced the front/back confusion rate and larger azimuthal movements ($\pm32^\circ$) drastically reduce, or even eliminate, such confusion. When virtual sources are coupled with the head-tracking system, the average reversal rate is usually higher, and wider movements are needed to reduce it \cite{wightman_resolution_1999} as the virtual sources are less externalised. Furthermore, McLachlan et al. 2023 \cite{mclachlan_dynamic_2023} showed that the front-back confusion errors were significantly reduced with small yaw rotations ($\pm10^\circ$) when stimuli were presented via loudspeakers but not when sources were binaurally reproduced with individual-HRTF coupled with the head-tracking system. This result suggests that small head movements are not effective enough in reducing front-back errors when stimuli are presented over headphones. 
Finally, Begault and colleagues (2001) \cite{begault_direct_2001} found no effect of reverberation in the front-back errors reduction (from $59\%$ to $28\%$) due to the use of head movements when sounds are presented via binaural spatial audio.
Those findings are consistent with our results when subjects use the head movements. The LME model suggests that mid-range movements (ROM $\le30^\circ$) significantly reduce the quadrant error rate independently on the acoustic condition. The presence of reverberation affects the relation only in the case of smaller movements (ROM $\leq20^\circ$) which is probably due to different externalisation levels across acoustic conditions. 
The use of different localisation strategies did not impact the localisation performance. On this topic, recent research is quite contradictory. Past studies have shown that the effective use of the head movements has a positive impact on localisation performance \cite{kato_effect_2003}, in particular, to reduce the elevation bias \cite{perrett_effect_1997}, even when the stimulus duration is relatively short (over 100ms \cite{ macpherson_cue_2013}). 
However, other studies have highlighted the fact that when subjects were not instructed on a particular strategy, large differences in the use of head rotations are reported and the optimal use of the dynamic cues needs to be learned \cite{wightman_resolution_1999,grange_benefit_2016,brimijoin_contribution_2013}; especially, when the individual spectral cues are disrupted \cite{valzolgher_reaching_2020, mclachlan_dynamic_2023}. 
For example, Begault et al. (2001) showed no improvement in localisation performance with the presence of head-tracking with a speech of 3s when subjects are instructed to freely move their head during listening. In summary, the reduced improvement in localisation performance with head movements could be explained by the fact that the used stimulus was too short (1.6s), limiting the ability of and effective use of the self-generated auditory cues, especially in the case of spontaneous movements.

It is widely recognised that reverberation reduces the localisation ability, disrupting the interaural coherence \cite{zahorik_spatial_2021} and reducing the accuracy in the horizontal plane \cite{giguere_sound_1993}. It has been shown that in static listening the presence of reflections increases the relative perceptual weight of the ILD at the detriment of ITD, which is more dominant (from the perceptual point of view) in anechoic conditions \cite{ihlefeld_effect_2011,nguyen_effects_2017}. According to \cite{devore_accurate_2009}, our brain uses this perceptual re-weighting to support an accurate localisation level in different acoustic conditions.
However, natural listening is, in general, a multi-modal task in which motor and sensory (e.g., auditory and visual) signals are constantly integrated \cite{myers_sensorimotor_2020} and so other strategies can be adapted to cope with the complexity of the real world. Our perception is a process where the systematic relation between actions and the associated changes of the sensory inputs (also known as, sensorimotor contingencies) are constantly learned. This points out also an important aspect of our perception which is the fact that is an active process in which the use of actions in the environment is crucial. 
Moreover, such listening behaviours are shaped by the environment in which we live and so by the statistics of the natural auditory world \cite{parise_natural_2014, pavao_natural_2020}.
In this context, it is possible to argue that our brain developed a multisensory approach to support a robust perception, adapting the listening strategy to the acoustic properties of the environment. Head movements in fact have a beneficial role in different auditory conditions. Grange et al. (2016) \cite{grange_benefit_2016} have shown that head movements can improve speech intelligibility in acoustic environments with spatially distributed sources. An optimal head orientation is beneficial to better separate auditory sources and reduce the interference between the target and the masking signals. It is also shown that head movements are crucial to improving source externalisation when sources are presented over headphones \cite{brimijoin_contribution_2013,hendrickx_influence_2017}. In the same way, the use of movements might be a spontaneous and effective strategy to reduce the uncertainty of the binaural cues due to the presence of reverberation \cite{heijden_active_2018}. The perceptual benefit of this strategy can be explained by exploiting a commonly used model in computational neuroscience: the divisive normalisation model. In computational neuroscience, there is a strong acceptance that due to the modular design of our brain, there is a set of computational processes that are applied in a variety of contexts which support our perception. One of these canonical operations is the divisive normalisation \cite{carandini_normalization_2012} which is demonstrated to underly different perceptual and behavioural processes, such as contrast adaptation in vision \cite{carandini_we_2005}, multisensory integration \cite{ohshiro_normalization_2011}, etc. In this context, the influence of reverberation reduces the binaural cues due to the decrease in IACC. Therefore, we could hypothesise that our brain would be able to normalise the ILD to the current IACC level which could limit the detrimental effect of reverb and the use of movements could be a spontaneous strategy used to cope with such a different listening condition. 
\begin{figure}[h!]
     \centering   \includegraphics{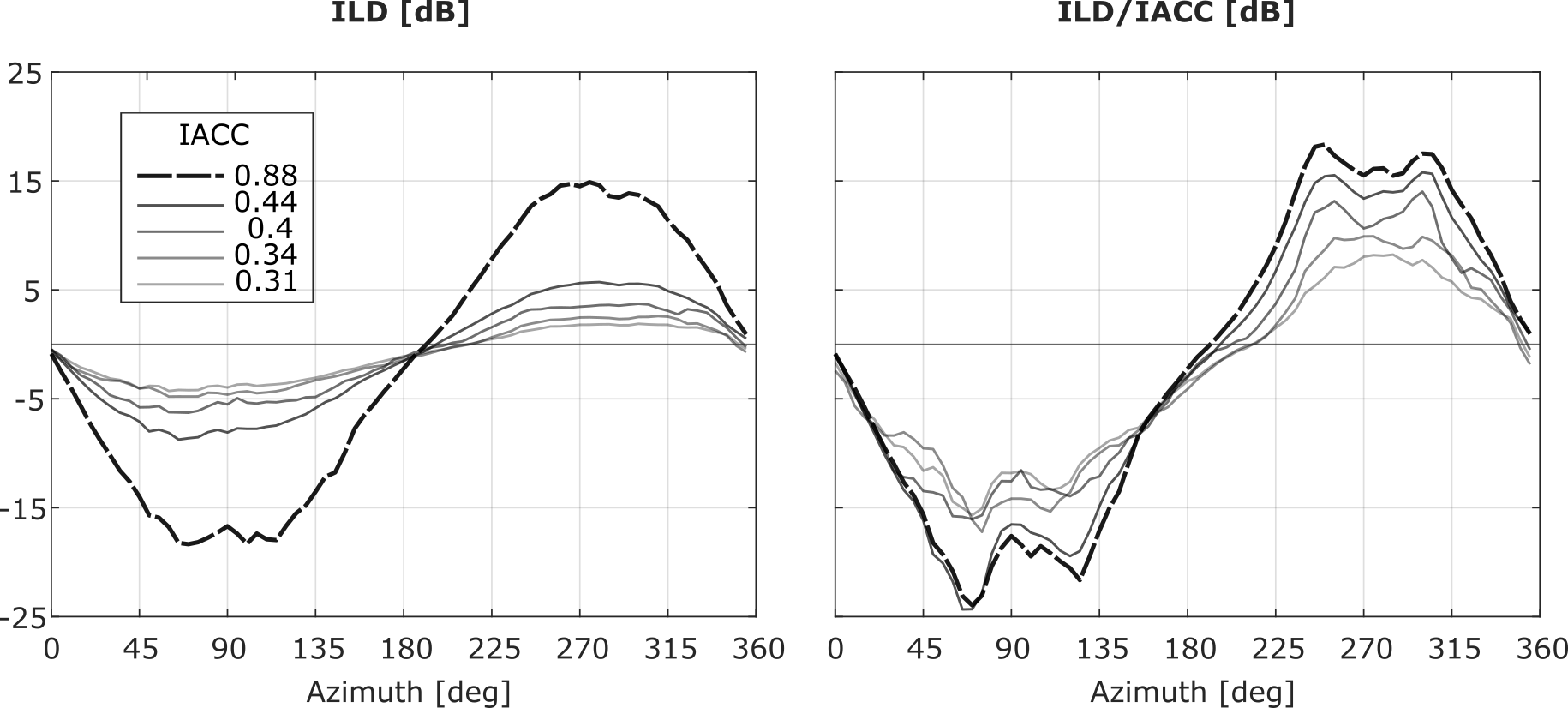}    \caption{\textbf{Effect on ILD of a normalization stage by means of IACC.} The panels show the same 
    ILD data shown in Fig.\ref{fig:aa_spatial_cues}B plotted as a function of azimuth locations in the reverberant conditions (grayscale lines, distance: d = [0.5m, 1m, 1.5m, 2m]) and a reference anechoic condition (dashed line, distance d = 0.5m) before (left) and after (right) a divisive normalization step where the ILD for each azimuth location has been divided by the associated IACC value (shown in Fig.\ref{fig:aa_spatial_cues}C). The inset in the left panel shows the average IACC value for the different conditions.}
    \label{fig:model}
\end{figure}
Fig.\ref{fig:model} shows the effect of normalisation on ILD cues when ILD on each azimuth location is divided by the actual IACC value, with different reverberation levels. As the IACC (i.e., distance) increases the ILD decreases, but when employing the normalisation model the disruptive effect of IACC is limited. As a consequence, the ILD in presence of reverb becomes more and more close to the values of the anechoic condition, where the IACC is close to 1.
The role of the increased movements with the presence of reverberation could be directed to this behavioural strategy: to reduce the uncertainty in the ILD due to the reverb, the brain could introduce the use of head movements to better separate the directional information of the direct and the diffuse signals and, therefore, reducing the detrimental effect of the reduced IACC.